\documentclass[aps,prb,twocolumn,groupedaddress]{revtex4}

\usepackage[final]{graphicx}
\usepackage{amssymb}

\makeatother

\newcommand{\MAE}{mixed--alkali effect}
\newcommand{\UNIT}[1]{\ensuremath{\mathrm{#1}}}
\newcommand{\VAR}[1]{\ensuremath{#1}}
\newcommand{\EXP}[1]{\ensuremath{\times{}10^{#1}}}

\newcommand{\CHEM}[1]{\ensuremath{\mathrm{#1}}}
\newcommand{\LI}{\CHEM{Li}}
\newcommand{\K}{\CHEM{K}}

\newcommand{\OCCI}{\ensuremath{o_i}}
\newcommand{\OCCILI}{\ensuremath{o_i^{\LI}}}
\newcommand{\OCCIK}{\ensuremath{o_i^{\K}}}
\newcommand{\OCCIM}{\ensuremath{o_i^{M}}}
\newcommand{\SPECI}{\ensuremath{s_i}}

\newcommand{\PILI}{\ensuremath{p_i^{\LI}}}
\newcommand{\PIK}{\ensuremath{p_i^{\K}}}
\newcommand{\PIM}{\ensuremath{p_i^{M}}}

\newcommand{\SOFT}{\ensuremath{S(t)}}
\newcommand{\TAURES}{\ensuremath{\tau_{res}}}
\newcommand{\TCH}{\ensuremath{t_{ch}}}
\newcommand{\TFREE}{\ensuremath{\tau_{free}}}

\newcommand{\PBACK}{\ensuremath{p_{back}}}

\makeatletter

\begin{document}


\title{Contributions to the mixed--alkali effect in molecular dynamics
       simulations of alkali silicate glasses}


\author{Heiko Lammert}
\email[]{hlammert@uni-muenster.de}
\author{Andreas Heuer}
\email[]{andheuer@uni-muenster.de}
\affiliation{Institute of Physical Chemistry and Sonderforschungsbereich 458, Corrensstr. 30, D-48149 M\"unster, Germany }


\date{\today}

\begin{abstract}
The mixed--alkali effect on the cation dynamics in silicate
glasses is analyzed via molecular dynamics simulations.
Observations suggest a description of the dynamics in terms of
stable sites mostly specific to one ionic species. As main
contributions to the mixed--alkali slowdown longer residence times
and an increased probability of correlated backjumps are
identified. The slowdown is related to the limited accessibility
of foreign sites. The mismatch experienced in a foreign site is
stronger and more retarding for the larger ions, the smaller ions
can be temporarily accommodated. Also correlations between unlike
as well as like cations are demonstrated that support cooperative
behavior.
\end{abstract}


\maketitle

\section{Introduction}

The disordered network structure of glasses provides a highly
complex environment for the transport of mobile ions. A basic
consensus has developed that the ionic motion consists of jumps
between distinct minima of the energy landscape
\cite{ingram:1987,baranowskii:1999}. Because of the strong Coulomb
interaction between the ions the effective energy landscape will
perpetually change with the motion of the ions. The dispersion in
the frequency dependence of ionic conductivity can be interpreted
\cite{funke:2000,cramer:2003} as a consequence of correlated
backward jumps forced by the interaction with other ions.

The complexity of the ion dynamics in glasses is highlighted
by the \MAE{} \cite{day:1976,tomandl:1985}:
In glasses containing two different alkali species, a deep minimum
in the mobility of the ions is observed.
Even small concentrations of a second alkali species effectively
slow down the total ion dynamics.

Several important structural properties of mixed--alkali (MA)
glasses are already known. EXAFS measurements \cite{greaves:1991}
have established that the cations retain the same specific
coordination environments as in single--alkali (SA) glasses. This
was also supported by reverse Monte Carlo calculations
\cite{Swenson:2001}.  Furthermore, NMR measurements
\cite{Yap:1995,gee:1996} have shown that the different alkali
cations are intimately mixed on the microscopic scale and do not
belong to phase segregated parts.

Specific information about the dynamics of cations in disordered
ion conductors can be obtained via Molecular Dynamics (MD)
simulations. For SA glasses it has been shown that regions of high
ionic mobility form channels \cite{jund:2002,sunyer:2003} that
arise statistically \cite{jund:2002,sunyer:2003,lammert:2004}.
Proof for some structure at the scale of these channels can also
be found in measurements of the dynamic structure factor
\cite{meyer:2002,meyer:2004}. One of the most important MD results
for MA glasses is the observation that ions only rarely jump into
sites vacated by the other species \cite{balasubramanian:1993} and
that the site characteristics of both species are very different
\cite{habasaki:1995,habasaki:1996,park:1999}.

Several models \cite{maass:1992,bunde:1994,Swenson:2001}, inspired
by the observations of site selectivity, feature distinct sets of
adapted sites for each alkali species.   Taking the site
selectivity as a starting point one may ask whether the dynamics
in the MA system of composition $(A_2O)_a (B_2O)_b
(SiO_2)_{1-a-b}$  can be viewed as the {\it independent}
superposition of two SA systems $(A_2O)_a (SiO_2)_{1-a}$ and
$(B_2O)_b (SiO_2)_{1-b}$, respectively. If this were true the
dramatic decrease of the dc conductivity in the MA system would be
related to the well-known observation that the alkali dynamics in
SA systems become much slower for lower alkali concentration and
one might just speak of a dilution effect.

Experimental results tell that reality is more complex. For
example, in recent measurements  of dc conductivities and tracer
diffusivities in Na--Rb--borate glasses the mobility of sodium and
rubidium is compared to that in pure Na--borates and Rb--borates
with identical alkali content  \cite{berkemeier:2004}. Despite a
significant decrease of the conductivity for the ternary glasses
the conductivity is still significantly larger than the sum of
those in the respective SA glasses. Furthermore, the experimental
data reveal asymmetric behavior with respect to the two alkali
species. A dramatic difference is visible when comparing the limit
where the smaller ion (here:Na) is the minority species as
compared to the opposite limit. It turns out that the small ion as
minority species is still relatively mobile and its diffusion
constant, as determined by tracer diffusion experiments, only
weakly depends on concentration. In contrast, in the other limit
the large ion becomes immobile. Similar observations have been
already made in earlier experiments; see e.g.
\cite{day:1976,ingram:1994}.

A simple reason for the apparent increase of mobility in mixed
alkali systems as compared to the sum of the individual glasses
may be given by the different network structures. Due to the
higher alkali content in the mixed--alkali system the network
structure is more discontinuous, giving rise to higher network
mobility and thus, possibly, to higher alkali mobility.

Beyond this effect it is discussed whether the adaption of the
individual sites may change with time. The rearrangement of free
volume has been put forward as a possible mechanism for the
readaption, based on measurements of the pressure dependence of
ionic conductivity \cite{bandaranayake:2002,ingram:2003}: A site
entered by a too large ion has to expand, while a site entered by
a smaller ion can shrink. If both happens at the same time in
close vicinity, an isochoric redistribution of site volume via
relaxations of the matrix could facilitate the readaption of both
sites. If this mechanism of \textit{matrix mediated coupling} were
indeed present, an important correlation between the unlike
cations would be a present in the MA system \cite{ingram:2003}.

Actually, the presence of the release of mismatches and thus the
readaption of sites in the glassy phase has been already
postulated in the Dynamic Structure Model \cite{bunde:1994}.
Others suppose that the adaption of a site is definitively fixed
during the glass transition \cite{Swenson:2001}. In a very
different approach it is assumed that a distribution of sites is
open in principle to all ions \cite{Kirchheim:2000,Hunt:1999}. In
this class of models, the competition for favorable sites is
governed by the different radii of the cation species.

The goal of our work is to elucidate the mechanism of ion dynamics
in mixed--alkali systems via MD simulations on a microscopic
level. The analysis is based on our method \cite{lammert:2003} to
identify individual ionic sites from the MD trajectories ; see
also \cite{habasaki:2004}. This procedure was also successfully
applied to simulations of phosphate glasses \cite{vogel:2004} and
rationalized in more general terms \cite{dyre:2003}. In this
investigation, the method is for the first time applied to a MA
system. From comparison of the dynamics in SA systems and MA
systems, three major aspects will be treated. First, the site
selectivity is quantified in detail. Second, the trajectories of
the alkali ions are characterized and compared with the typical
behavior in SA systems. Third,  dynamic correlations between like
and unlike cations are analyzed.

\section{Technical Aspects}

We performed molecular dynamics simulations of alkali silicate
glasses
$x(\mathrm{K_2O})\cdot{}(1-x)(\mathrm{Li_2O})\cdot{}2(\mathrm{SiO_2})$
with $x=0.0,0.5,1.0$. We use  a modified version of the software
MOLDY \cite{refson:2000}, using Buckingham--type pair potentials
developed by Habasaki \textit{et al.} \cite{habasaki:1992}, that
are well tested for comparable systems
\cite{habasaki:1998,banhatti:2001,heuer:2002}. The simulations
were done in the NVT--ensemble, with systems of 1215 particles,
i.e. 270 alkali ions. The size of the simulation boxes was set
according to experimental densities \cite{glassprops}, resulting
in $V(x=0):V(x=0.5):V(x=1) = 1:1.03:1.06$.  In
\cite{habasaki:1992} the experimental densities for the two SA
systems could be reproduced from NpT-simulations at identical
pressure. In contrast, we obtain differing pressures
in our NVT runs with these experimental densities.
At $T=3000$K, where equilibrium runs are easily
possible, the pressure of the two limiting systems for our
simulations are $p(x=0) \approx 2.5$ GPa and $p(x=1) \approx 4.7$ GPa.
Yet the pressure of the MA system is roughly the average
of the two limiting SA systems, with a deviation of $0.4 \pm 0.2$ GPa.
This relation does not change at
different temperatures. The additive behavior supports the idea that the
observed slowdown in the MA system (see below) is not generated by our 
choice of densities.

The systems were prepared at high temperatures and cooled down in
steps. Here we present data from runs at $850 \mathrm{K}$, which
is well below the glass transition. The alkali subsystem was
equilibrated at this temperature for $40 \UNIT{ns}$. Subsequently
trajectories were generated over $40 \UNIT{ns}$ for the SA
glasses, and over $160 \UNIT{ns}$ for the MA system. The
determination of cation sites based on cation trajectories is
described in detail in \cite{lammert:2003}.

\section{Results}

\subsection{General dynamics}

\begin{figure}
\includegraphics[width=8.6cm,clip]{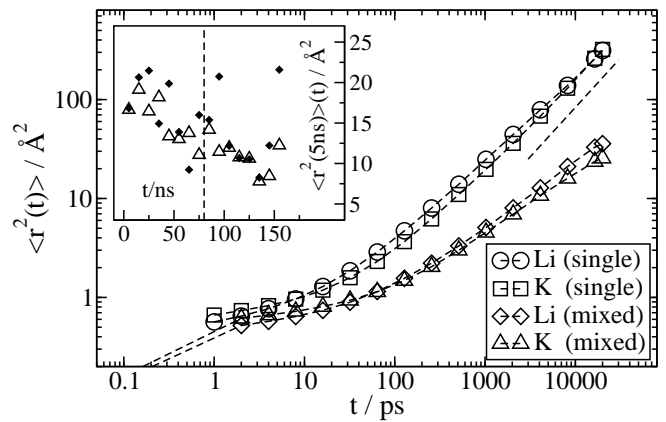}
\caption{\label{fig_diff}Mean square displacements of alkali ions
in single and mixed--alkali systems. Inset: Time dependence of
$<r^2(5 \UNIT{ns})>$ in the mixed--alkali system.  The broken line
at $80 \UNIT{ns}$ indicates that for this work only information
from the second part of our data is taken if not mentioned
otherwise.
 }
\end{figure}

Simulations below the computer glass transition temperature have
to be carefully conducted due to expected aging effects of the
network. Furthermore one may expect that due to this effect also
the lithium dynamics will somewhat change with time. After an
initial simulation period of $40 \UNIT{ns}$ we have calculated the
mean square displacement during time intervals $[t_0,t_0 + 5
\UNIT{ns}]$ where different starting times $t_0$ during our
production run of  160 \UNIT{ns} have been chosen; see inset of
Fig. \ref{fig_diff}. During the first $80\UNIT{ns}$ a significant
slowdown is present, which is less visible for the second $80
\UNIT{ns}$. Thus one finds a dynamic signature of network aging
effects. Of course, subsequent aging effects of the network will
still be present, but probably on longer time scales. The
subsequent analysis is mainly performed for the second half of our
production run.  In any event, all general conclusions drawn in
this work could already be obtained from analyzing just the first
half of the production run.

Figure \ref{fig_diff} shows the mean square displacements of the
alkali cations in both the SA and MA glasses. In the long time
limit the ionic transport becomes diffusive, corresponding to a
slope of one in the double logarithmic plot. In the SA glasses
this is reached around $t = 10 \UNIT{ns}$.  The diffusivities at
$850 \UNIT{K}$ in the SA systems are $D_{Li} = 2.9 \EXP{-7}
\UNIT{cm^2s^{-1}}$ and $D_{K} = 2.6 \EXP{-7} \UNIT{cm^2s^{-1}}$.
The curves for lithium and potassium in the MA system do not
attain a slope of one during the time interval shown. This
observation is in agreement with earlier findings in SA glasses,
where we noted that the diffusive regime is reached only for
larger values of the mean square displacement in systems with
lower alkali content \cite{lammert:2003}. Under the assumption of
time-temperature superposition and the ability to reach the linear
regime at higher temperatures we may estimate the diffusion
constants as  $D_{Li} = 2.8 \EXP{-8} \UNIT{cm^2s^{-1}}$ and $D_{K}
= 1.9 \EXP{-8} \UNIT{cm^2s^{-1}}$. The slowdown of the
diffusivities as compared to the single alkali systems then
amounts to a factor of $\approx{} 10$ for lithium and $\approx{}
14$ for potassium.

\subsection{Properties of ion sites}

By mapping the trajectories of the cations onto the sites obtained
from our analysis, jumps between the sites can be identified. As
in our previous work, we registered a jump if an ion leaves a site
and moves into a different one. The duration of an ion's residence
in a site, denoted \TAURES{}, is defined by the time between its
jump into the site and the subsequent jump out of it. When an ion
leaves a site and returns without reaching a different site, the
residence is not interrupted. In total we find 288 sites. Thus,
there are ca. 8\% more sites than ions. This small excess number
is close to the result for the SA system \cite{lammert:2003}.

\begin{figure}
\includegraphics[width=8.6cm,clip]{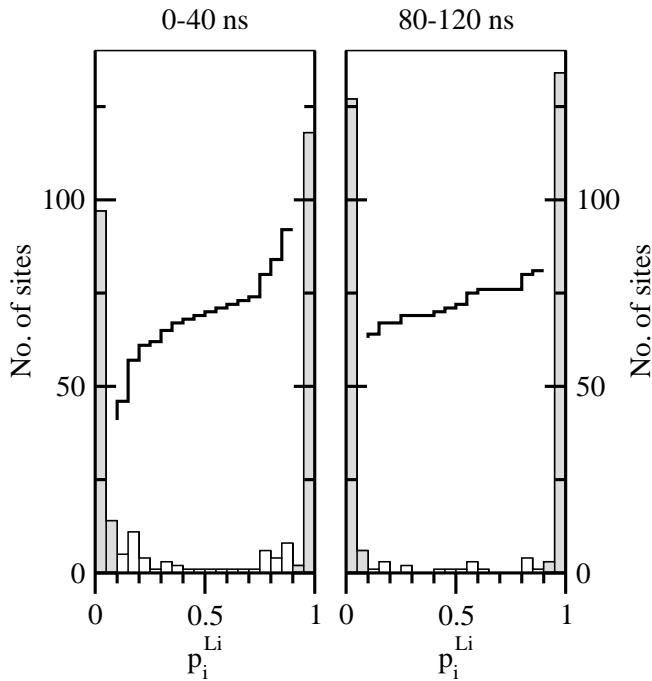}
\caption{\label{fig_siteada} Distribution of \PILI, the fraction
of time a site is occupied by lithium cations. This is a measure
of specificity (see text). The shaded areas correspond to adapted
sites, characterized by specificities $ \ge 0.9$. The bold black
lines give the cumulative sum over the number of mixed sites,
shifted in vertical direction for convenience. }
\end{figure}

>From the residence data it is also possible to calculate the
average occupation \OCCI{} for each site $i$. In the MA system we
distinguish the contributions \OCCIM of the different cations
$M=\{\LI,\K\}$. The specificity of a site can be quantified by
$\PILI = \OCCILI/(\OCCILI + \OCCIK)$ and $\PIK = 1-\PILI$, the
probabilities that an occupying ion is lithium or potassium,
respectively. One can define the specificity
\SPECI{}=max($\PILI,\PIK$) as the larger of both \PIM. The
distribution of \PILI \, is shown in Fig. \ref{fig_siteada}, for
the interval $80-120 \UNIT{ns}$ and also for $0-40 \UNIT{ns}$ for
comparison. The qualitative result is identical. At both times
most sites have an $\SPECI \ge 0.9$, signalling a high specificity
for one predominant ion. For the subsequent analysis these sites
will be termed {\it adapted} sites or, alternatively, lithium or
potassium sites. Those with lower specificity shall be called {\it
mixed} sites. The regions corresponding to adapted sites are
shaded in Fig. \ref{fig_siteada}. Also shown are the cumulative
sums over the number of mixed sites, shifted to a convenient
vertical position. They clearly show that the number of mixed
sites has decreased at the later time. Especially mixed sites with
a $\SPECI$ already close to 0.9 have vanished, with a
corresponding increase in the number of adapted sites.
At $80-120 \UNIT{ns}$, 94\% (270) of all sites 
have a $\SPECI \ge 0.9$, compared to 82\% (231) at $0-40
\UNIT{ns}$.
25\% of all sites have been 
exclusively populated by lithium cations, and 19\% have only been
visited by potassium cations between $80-120 \UNIT{ns}$. The
values for $0-40 \UNIT{ns}$ are slightly lower again, being 23\%
and 17\%, respectively. The differences between the time slices
illustrate the effects of aging on our data.

The observed dependence of site specificity on simulation time
rationalizes the observed time dependence of alkali mobility in
the MA system. For later times the site specificity is more
pronounced and thus the MA effect, expressed by a slowing down of
the alkali dynamics, increases.

\begin{figure}
\includegraphics[width=8.6cm,clip]{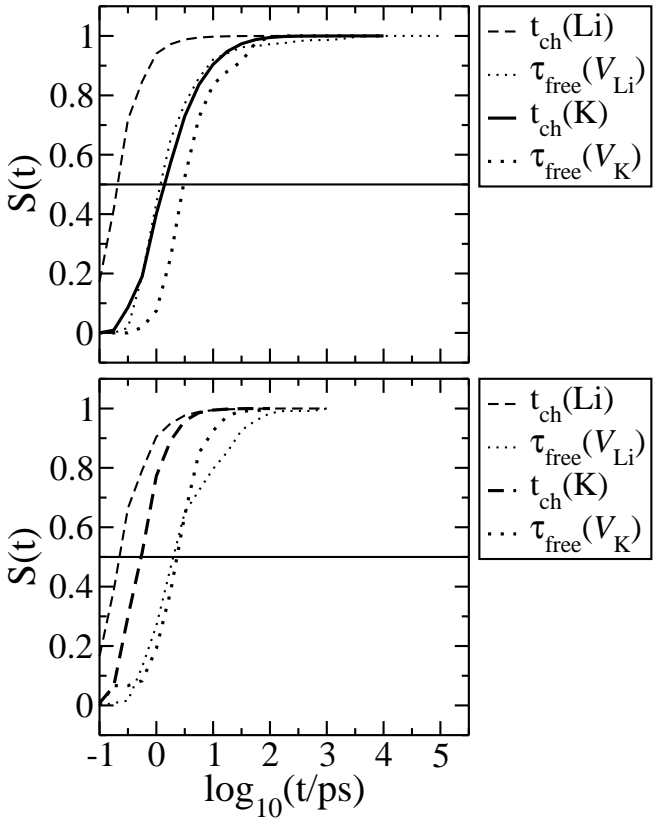}
\caption{\label{fig_times_short}Distributions of  duration of
jumps \TCH{} and of \TFREE{}, the time a site stays unoccupied.
Top: single--alkali systems; Bottom: mixed--alkali system. From
the horizontal line with value 0.5 the median can be read off. }
\end{figure}

\begin{figure}
\includegraphics[width=8.6cm,clip]{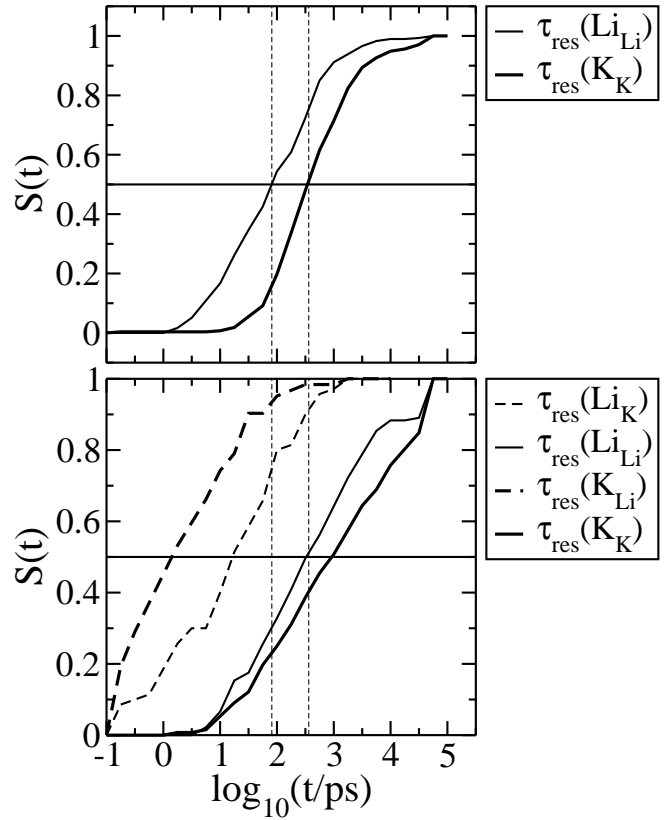}
\caption{\label{fig_times}Distributions of residence times
\TAURES{}. Top: single--alkali systems; Bottom: mixed--alkali
system. }
\end{figure}

\subsection{Site properties}

In previous work on a SA system we have shown that the time scale
it takes for an ion to hop from one equilibrium position to a
nearby position is very short (less than 1 ps for lithium)
\cite{kunow:2005}. This time scale is denoted \TCH{}. Furthermore,
it turned out that on average a site was filled very soon after it
was left by an ion. This is quantified by \TFREE{}. The low values
reflect the fact that the number of free sites is very small. We
have determined the cumulative distribution $S(t)$ of both
quantities for both the SA and the MA system.  Thus, $S(t)$
expresses the probability, that a value $\leq t$ is found. The
results are shown in Fig. \ref{fig_times_short}.

For both systems it turns out that both times belong to the
ps-time scale. Of course, the short transition times  \TCH{}
directly imply that ionic transitions are exclusively between
adjacent sites. There is no long-range ionic motion through the
network before entering a new site. Interestingly, in the SA
system the potassium dynamics between two sites is nearly a factor
of 10 slower than the lithium dynamics. This difference alone
cannot be explained only by the mass ratio of potassium and
lithium. Actually, it becomes smaller in the MA system. It may
come as a surprise that the time \TFREE{} during which a site is
vacant is not much longer than the transition time \TCH{} between
adjacent sites. This reflects strongly cooperative dynamics, as
stressed in previous simulation work
\cite{park:1999,habasaki:2004a}. Interestingly, the distributions
are significantly broader for the MA system. This is a first hint
that the dynamics is more heterogeneous in the MA system.

In the next step we calculate for every site its residence time,
averaged over all ionic visits of this site. In the MA case we
distinguish whether the visit is in a matched site or a mismatched
site. Thus, $\TAURES(\LI_{\mathrm K})$ corresponds to the waiting
time of a lithium ion in a potassium site (characterized by $\PILI
< 0.1$). In Fig. \ref{fig_times} we show the corresponding
cumulative distributions \SOFT{}. Comparing the data from the SA
systems shown in the top part with those of the MA system
($\TAURES(\LI_{\LI})$ and $\TAURES(\K_{\K})$, respectively) it
turns out that the distribution is significantly broader in the
latter case, in particular for long times. This means that the
dynamics is more heterogeneous in the MA system. Actually, from
previous work we know that the width of the distribution of
residence times in the SA case does not depend on concentration
\cite{lammert:2003}. Thus, the increasing relevance of dynamic
heterogeneities in the MA case is our first observation, for which
the dynamics displays qualitatively new features in MA systems as
compared to SA systems. Interestingly, the median values of the
distributions of residence times increase by a factor of only ca.
three in the MA system. Thus the increase in residence times is
not sufficient to explain the total extent of the MAE (factor
10-14, see above).

>From Fig. \ref{fig_diff} one can see that for the SA system the
crossover to diffusive dynamics occurs on time scales somewhat
longer than 100 \UNIT{ps}. Here we define crossover such that the
extrapolation from the long-time diffusive behavior deviates by a
factor of 2 from the actual mean square displacement. This time
scale is close to the median of the waiting times. Thus one may
conclude that for the SA system the dynamics of ions from the
slower half of the sites is not important for understanding the
long-range transport. In contrast, for the MA case this crossover
time can be estimated to be significantly beyond 10 \UNIT{ns}.
This implies that nearly the whole range of residence times is
relevant for diffusive transport.

So far we have analyzed the behavior of matched events, i.e.
adapted sites hosting the appropriate ion. Analyzing the waiting
times for mismatched events we find, in particular for potassium,
a strong reduction of the waiting times. This asymmetry in the
behavior of lithium and potassium is also mirrored by the lower
number of exclusive potassium sites mentioned in the discussion of
Fig. \ref{fig_siteada}. Qualitatively, this may be related to the
fact that it is difficult for a potassium ion to enter a smaller
lithium site. A more quantitative version of this argument will be
presented below.

\begin{figure}
\includegraphics[width=7.1cm,clip]{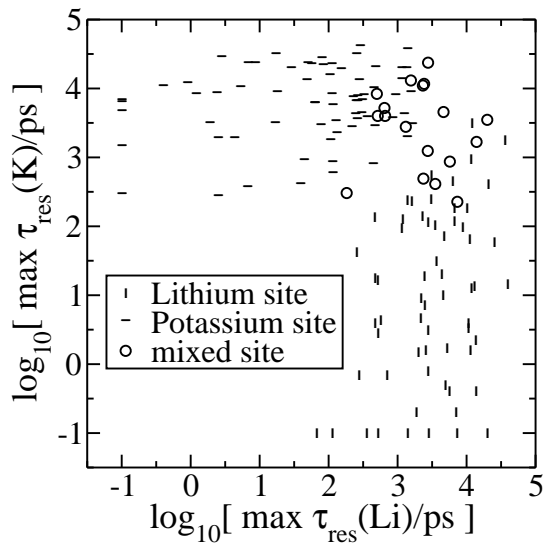}
\caption{\label{fig_maxtaukli} Maximum residence times in sites
visited by lithium and potassium ions during the simulation. For
each site, the value for potassium is plotted versus the value for
lithium. Adapted sites and mixed sites are distinguished by their
relative occupation.}
\end{figure}

Although residences for mismatched events are on average
significantly shorter than the residences found in matching
events, the ranges covered by the distributions in Fig.
\ref{fig_times} are overlapping. The duration of mismatching
events can thus reach time scales normally present in well adapted
sites. This possibility is explored in Fig. \ref{fig_maxtaukli}.
For each site that has been visited by both lithium and potassium
ions, the maximum residence time of a potassium ion is plotted
against the maximum residence time of a lithium ion in the same
site. Values for lithium sites, defined as above by $\PILI{}>0.9$,
are noted as vertical lines, potassium sites as horizontal lines.
Mixed sites are shown as circles.

The few mixed sites allow long maximum residence times for both
species of ions, with values of at least 100\UNIT{ps}, but
typically reaching several \UNIT{ns} for at least one species. All
the adapted sites show residence times in this range for ions of
the favored kind. The maximum residence time of a mismatched ion
can be very short. But surprisingly, the possible residence times
of mismatching events too cover the whole range of values up to
comparable maximum time scales for both species, despite the high
specificity of the respective sites. Even some strongly adapted
sites seem thus to have the potential to accommodate ions of both
species for a long time.

\begin{figure}
\includegraphics[width=7.1cm,clip]{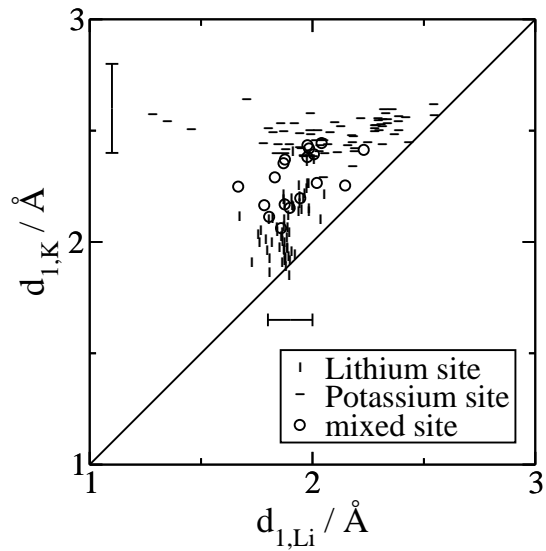}
\caption{\label{fig_rox1kli} The mean distance  from a site to the
closest oxygen during potassium residences, i.e. $d_{1,K}$, is
plotted versus the value found during lithium residences in the
same site, i.e. $d_{1,Li}$. The bars give the typical ranges of
values for sites visited by only one species. }
\end{figure}

As a further step it is instructive to analyze the structural
aspects of site specificity. The oxygen coordination environment
of lithium and potassium sites is very different. One finds 4.5
oxygens up to $r\leq{}2.87 \UNIT{\AA{}}$ for a lithium site and 10
at $r \leq{} 4.01 \UNIT{\AA{}} $ for a potassium site. The
distances correspond to the first minima of the distribution
function. A different way to characterize the differences is the
observation that around a lithium ion, on average, three
coordinating oxygen atoms are closer than 2.3 \UNIT{\AA{}}, while
on average all coordinating oxygen atoms are further apart around
a potassium ion. We determined for each site the mean distances
$d_{i,M} (i=1,2,3,4; M = \mathrm{Li,K})$ between the center of the site and
the i-th closest oxygen atom whenever this site was either
occupied by a lithium or a potassium ion. In particular $d_{1,M}$
is an extremely sensitive quantity to the local structure of a
site. In Fig. \ref{fig_rox1kli} $d_{1,K}$ is plotted vs.
$d_{1,Li}$ for all sites that have been visited by ions of both
species. Again, data for lithium sites, for potassium sites and
for mixed sites are distinguished by different symbols. The ranges
of typical values for pure sites which are exclusively occupied by
lithium or potassium, respectively, are given by the bars parallel
to the respective axis. The bars cover more than 90 \% of the
values. In Tab. \ref{median} we have for all combinations between
ions and sites compiled the intervals for $d_{1,M}$  which apply
to 68\% of all sites.
 \begin{table}
  \begin{center}
  \begin{tabular}[t]{|c|c|c|c|c|}\hline
       \multicolumn{1}{|c|}{\parbox[t]{1.0cm}{}}&
       \multicolumn{1}{|c|}{\parbox[t]{1.0cm}{Li site}}&
       \multicolumn{1}{c|}{\parbox[t]{1.0cm}{K site}} &
       \multicolumn{1}{c|}{\parbox[t]{1.0cm}{pure \\ Li site}} &
        \multicolumn{1}{c|}{\parbox[t]{1.0cm}{pure \\ K site}}\tabularnewline[1mm]
       \hline
       Li  ion & [1.83,1.94]\AA & [1.87,2.33] \AA & [1.84,1.93]\AA & - \tabularnewline[1mm] \hline
     K ion  & [1.93,2.20]\AA & [2.43,2.60]\AA & - & [2.47,2.64]\AA  \tabularnewline[1mm] \hline
           \end{tabular}
  \caption{\label{median} The intervals for the distance of the center of a site
  to the nearest oxygen where 68\% of all cases are included. Only adapted sites are taken into account.}
  \end{center}
\end{table}

Several interesting conclusions can be drawn about the distance to
the next oxygen atom from Fig. \ref{fig_rox1kli} and the
corresponding data of Tab. \ref{median}. First we discuss the
properties of adapted sites. (1) Except for a single site one has
$d_{1,Li} < d_{1,K}$. This is a natural consequence of the
different radii of the two ion species. (2) The median of
$d_{1,Li}$ and even the total distribution is very similar when
comparing pure sites with adapted sites, which at least once have
hosted a potassium site. Thus the occasional occupation by
potassium ions does not change the local structure of a lithium
site.  (3) The reverse is not true. Pure potassium sites have on
average larger values of $d_{1,K}$ than adapted sites which at
least once have hosted an lithium ion.  Thus lithium ions can only
visit those potassium sites with not too large values of
$d_{1,K}$.  (4) Lithium ions visiting a potassium site
significantly attract the nearest oxygen. There is, however, only
a partial adaption since the median of $d_{1,Li}$ is still 0.20
\AA \, larger than the median for lithium sites.  (5) In contrast,
lithium sites only weakly adapt if entered by potassium ion. This
is plausible from a chemical point of view because it is more
difficult to compress the oxygen coordination shell rather than
expanding it.

What can we learn from the properties of the mixed sites? A priori
the mixed sites may have two different origins. First, they may
reflect sites which during the first part of the 40ns were either
lithium or potassium sites and in the second part changed their
identity. This would give rise to a mixed average occupation.
Second, they are never potassium or lithium sites but just have a
structure which allows both ions to enter this site.  Whereas the
results in Fig. \ref{fig_maxtaukli} do not allow us to distinguish
between both scenarios the data in Fig. \ref{fig_rox1kli} clearly
demonstrates that the second scenario is valid. If during some
time interval the mixed sites would have acted as potassium sites
the values of $d_{1,K}$ should be in the range of the values for
potassium sites. Clearly, this is not the case. Thus a relevant
number of site transitions from lithium to potassium type or vice
versa cannot be seen.

\begin{figure}
\includegraphics[width=7.1cm,clip]{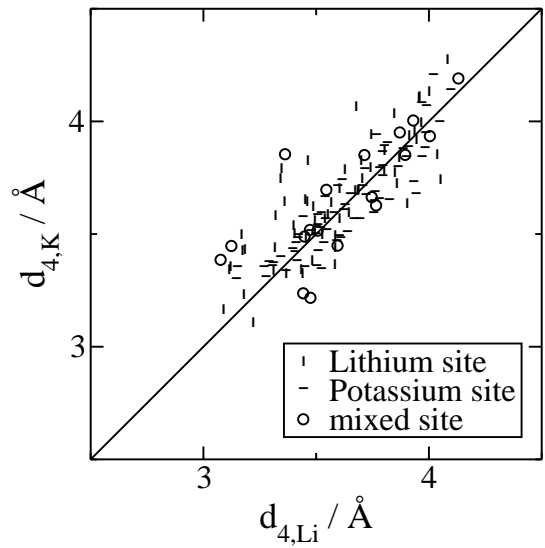}
\caption{\label{fig_rox4kli} Mean distances $d_{4,K}$ and
$d_{4,Li}$ from mixed and adapted sites  to the fourth oxygen
neighbor. }
\end{figure}

The omitted corresponding plots for the second and third neighbors
show gradually weaker displacements. The distance to the fourth
oxygen neighbor, given in Fig. \ref{fig_rox4kli}, finally does not
show any effect of a change in the occupying species. The reaction
of the network around a site to the nature of a visiting cation is
thus very limited in range.

>From these observations we would particularly like to stress that
lithium and potassium sites only differ in the behavior of the
first three oxygen atoms and that the properties of lithium sites
do not change significantly, even if they are visited by potassium
ions. This directly rationalizes the behavior, reported above,
that the residence time of potassium in lithium sites is very
short.

\subsection{Nearest-neighbor jumps}

\begin{figure}
\includegraphics[width=8.6cm,clip]{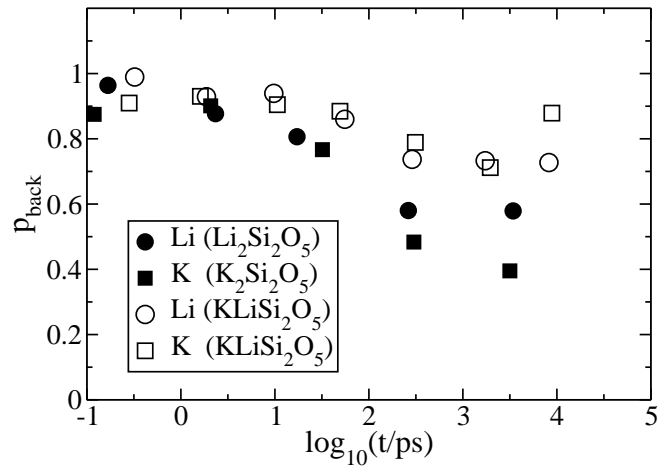}
\caption{\label{fig_pback}Probability of direct backjumps A
$\rightarrow$ B $\rightarrow$ C  plotted against the average
residence time for site \VAR{B}}.
\end{figure}

\begin{figure}
\includegraphics[width=8.6cm,clip]{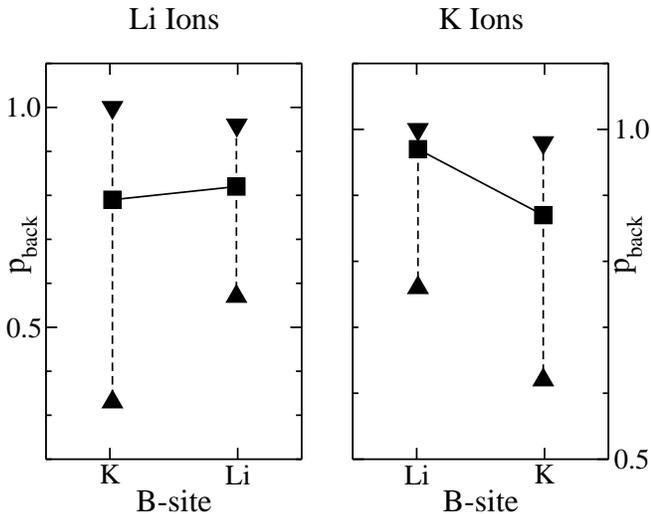}
\caption{\label{fig_pranges} Comparision of \PBACK{} out of
matched and mismatched sites in the mixed--alkali system, for both
lithium and potassium ions. The squares give the median of the
distribution of one-site values, the triangles delimit the range
including 34\% of all sites both below and above the median. }
\end{figure}

The local cation neighborhood of a cation determines, of course,
the properties of nearest neighbor (nn) jumps.  An important
observable is the coordination number in the nearest neighbor (nn)
shell, based on the partial pair correlation functions. Basically
one counts the number of neighboring ions until the first minimum
of the respective partial pair correlation functions. For Li-Li
pairs we obtain 3.6 for K-K pairs 4.1 and for Li-K pairs 3.0. This
clearly shows that, on the one hand, there is a slight tendency
for clustering of like ions and, on the other hand, potassium ions
have more neighbors than lithium ions. The latter point is a
direct consequence of the fact that potassium has a larger ionic
radius than lithium and thus has a larger nn sphere. It directly
translates into the observation that also the number of nn sites
is larger for potassium (8.1) than for lithium (7.1).

Of direct relevance for the dynamics are the number of nn sites
which are accessible by a hopping process. It turns out that only
ca. 60\% and 40\% of all nn sites were used as destinations for
jumps by lithium and potassium ions, respectively. For the slowest
sites, not all suitable neighbors may have been explored during
the analyzed interval, especially in the case of potassium. But
even considering this possibility, a significant fraction of nn
sites must remain where the saddles are too large for a hop to
occur. In the next step one may analyze the nature of those
neighbors actually accessed
during the simulation. It turns out that 70\% of them are 
matching for the jumping ion. This value is significantly larger than the
overall fraction of matching sites
(ca. 47\% for both species; see above). 
Matching sites are thus clearly favored as jump destinations.

Beyond the residence times the long-time dynamics is strongly
influenced by the presence of correlated forward--backward jumps.
To quantify this effect, we consider sequences of two jumps A
$\rightarrow$ B $\rightarrow$ C performed by an ion. We define
\PBACK{} as \mbox{$p(C=A)$}, the probability that an ion in a site
\VAR{B} jumps back into the same site \VAR{A} from where it
reached \VAR{B}. In Fig. \ref{fig_pback}, \PBACK{} is plotted
against the average residence time of an ion in site \VAR{B},
$\TAURES(M_B)$.

In all systems, high probabilities are found for backjumps through
sites \VAR{B} with short residence times, with values approaching
$90 \%$. \PBACK{} falls off for longer residences, down to $50 \%
$ and $40 \%$ respectively for the lithium and potassium SA
systems. These values are still high compared to the statistical
probability of $15 \%$ corresponding to the average number of
seven available neighbors in both cases. In the MA system, the
curves level off to even higher values of \PBACK{} for long times,
$65 \%$ and $70 \%$ for lithium and potassium. Generally, the
probability of backjumps for any given residence time is higher
than in the SA systems. The difference between the SA and the MA
systems directly translates into the longer range of subdiffusive
behavior (see above) and thus to a further decrease of the
diffusion constant for the MA systems. This means that beyond the
(minor) effect of longer waiting times it is to a large extent the
increase of forward--backward correlations which gives rise to the
slowdown in the MA system.

For a closer understanding of the backjump characteristics we have
studied \PBACK{} for different sites. From Fig. \ref{fig_pback} it
is evident that short residence times are generally related to
high backjump probabilities. This expectation is directly checked
in Fig. \ref{fig_pranges}. Therein the observed ranges for
\PBACK{} out of lithium and potassium sites are opposed, for both
species of ions.  For potassium, the expected strong rise in
\PBACK{} is found if \VAR{B} is a mismatched site. But the
backjump probability of lithium ions is only slightly affected by
the nature of the site \VAR{B}. This underlines our conclusions
about the hopping properties of potassium ions. Whenever a
potassium ion enters a mismatched site it can hardly enter because
it is too difficult to create the necessary volume and immediately
jumps back to the original site.

\subsection{Dynamic correlation effects}

\begin{figure}
\includegraphics[width=8.6cm,clip]{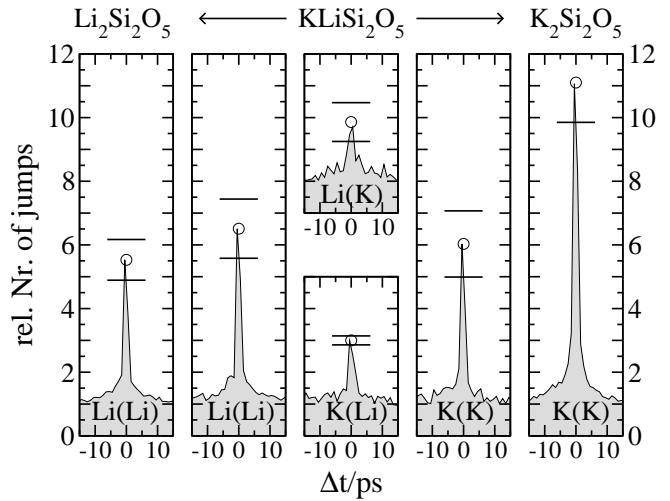}
\caption{\label{fig_correl}Relative number of jumps from the neighboring sites
a time $\Delta t$ from a jump out of a given central site. In brackets the
ion jumping at the center, in front the species jumping in response.
The distribution is the average over those for any given central site.
}
\end{figure}

\begin{figure}
\includegraphics[width=4.0cm,clip]{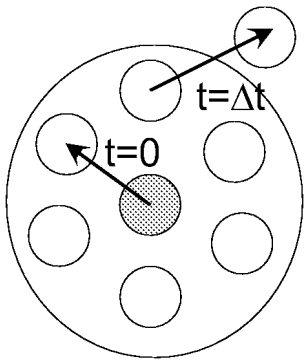}
\caption{\label{scheme} Schematic of the type of correlated jump
processes, considered in our analysis where four distinct sites as
start and end points are required. The central site is shaded.}
\end{figure}

\begin{figure}
\includegraphics[width=8.6cm,clip]{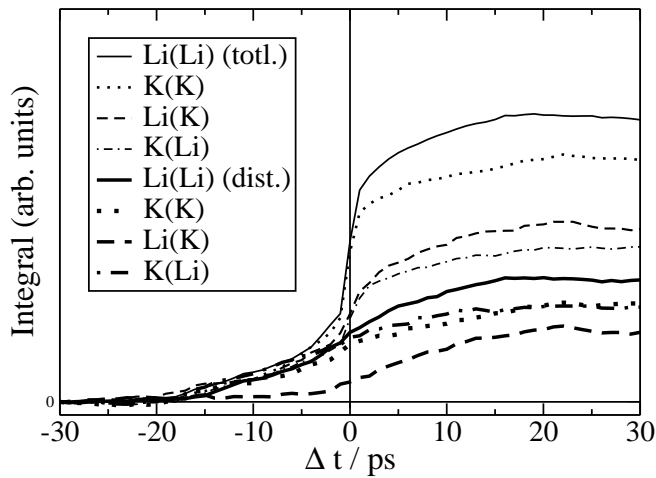}
\caption{\label{fig_intcorr} Cumulative amount of additional jumps
correlated with a jump at $t = 0$, for the mixed--alkali system.
The importance of correlated jumps between four distinct sites is
compared to the total extent of correlations including sequential
movement as shown in Fig. \ref{fig_correl}. }
\end{figure}

So far we have analyzed the dynamic properties on a {\it
single-particle} level. As discussed in the introduction an
influence of {\it multi-particle} correlations is discussed in
literature as a possible mechanism to enable dynamic processes in
MA systems. For example Habasaki demonstrated an additional
slowdown of the lithium dynamics, if the potassium ions in a MA
system were completely fixed \cite{habasaki:2004a}. The inferred
correlations between jumps of different ions should be directly
observable as an altered probability for a second jump shortly
before or after one jump has happened. For each site, we have
determined the number of jumps originating from one of its nn
sites in dependence on the time difference $\Delta t$ to a jump
from the central site. Jumps by the same ion were ignored. The
resulting distribution of jump probabilities was normalized and
averaged over all sites. The value of one corresponds to the
number of jumps from the neighbors happening without the influence
of a jump from the central site, i.e. the behavior for long
$|\Delta t|$. To gain an estimate of the scale of fluctuations in
the results, this analysis was done separately for four subsequent
pieces of the simulation of $10 \UNIT{ns}$ duration each. The
distributions shown in Fig. \ref{fig_correl} are the average over
these four partial distributions. The error bars correspond to the
standard deviation of the peak heights. The peaks show an increase
in the number of nearby jumps by a factor of $5.5$ and $11$ for
the lithium and potassium SA systems, respectively. For the
correlations among like ions in the MA system, a factor of $6 -
6.5$ is found for both lithium and potassium. Correlations among
jumps by unlike ions are weaker, but also present. The values are
$3$ for the lithium neighbors of a potassium ion, and also for
potassium ions surrounding a lithium. In all cases, the maximum
probability for correlated jumps is found at the time of the first
jump, $\Delta t = 0$, and the peak is symmetrical. Careful
inspection reveals that the peak consists of a narrow peak with
half width close to 1 ps and a broad peak with with half width
close to 5 ps. The latter peak will be discussed below.

The most simple scenario compatible with these findings is that a
second ion jumps from a neighbor site into the site just vacated
by the jump of the first ion. Indeed these correlations are
reflected by the sharp peaks. But more complex correlations are
also conceivable, which can generally be separated from the former
kind by the additional condition that the two jumps use four
distinct sites as start and end points; see Fig. \ref{scheme} for
a sketch.

The remaining correlations are responsible for the broad peaks in
Fig. \ref{fig_correl}. For a quantitative analysis the maximum
height is not suitable. Rather we show the integrals $\int_{-
\infty}^{\Delta t} \, d \tau (peak(\tau)-1)$ over the peaks; see
Fig. \ref{fig_intcorr}. On the one hand, we integrate over the
total peaks and, on the other hand, only over the remaining
correlations if we use the additional four-site condition
introduced above.

The total extent of four--site correlations given by the thicker
lines is significant in all cases, amounting to ca. 50\% of the
total. To characterize the observed correlations, we determined
the relative directions of the central jump and the correlated
ones, given by the cosine of the enclosed angle. The directions of
jumps using a common site are strongly correlated in all cases, as
it is geometrically required. The jumps involved in four--site
correlations are instead preferentially antiparallel in the SA
systems. In the MA system, no correlation is discernible for the
four--site correlations. Interestingly, the curve labelled Li(K),
i.e. the probability of a lithium jump to follow a potassium jump,
displays an asymmetry since it mainly increases for $ \Delta{} t
> 0$. Accordingly, the curve K(Li)shows the opposite asymmetry.
Both observations imply that it is more likely that a jump of a
potassium ion may trigger a jump of a lithium ion than vice versa.
This observation is consistent our previous results that potassium
sites are better suited for lithium ions than vice versa. A more
detailed analysis, however, is necessary to further clarify the
origin of this asymmetry.

In this analysis of correlations, all sites are taken into account
with equal weight. When the sites are weighted according to the
number of jumps taking place, the general effects remain, but
their strength is greatly reduced. One can thus conclude that
dynamic correlation effects are particularly important in the
vicinity of slow sites.

\section{Discussion}

The results favor an explanation of the \MAE{} based on ionic
sites that are specific for the different alkali species. The
strongly bimodal distribution of occupation probabilities \PIM{}
shows that $90 \%$ of the ions reside in sites with a strong
preference for one kind of ion. We have shown before that the
location of the ionic sites is stable on the time scale of ionic
transport \cite{Heuer:2004}. It appears thus justified to treat
the preference of a site for one species of ions as fixed over the
course of the simulations. Possible effects of rare readaptions
would still be incorporated via the treatment of "mixed" sites.

Trying to quantify the \MAE{} in terms of properties of
single sites or ions, the most direct contribution is observed in
the distribution of residence times. The typical duration of
residences in the predominant specific sites in the MA system is
increased compared to the SA glasses. But this effect can account
for only a small part of the observed \MAE{}.

The increased probability for correlated forward--backward jumps
will also add to the slowdown. As already in the SA systems the
effect is strongest for sites with low residence times. Among them
will be mismatching jump events, which generally have shorter
\TAURES{} than matching ones. But only for potassium ions this
tendency translates into a further increase of \PBACK{}. The
backjump probability of lithium is hardly affected by the adaption
of the \VAR{B} site.

The maximum residence times and the oxygen coordination distances
show that even strongly adapted sites can temporarily accommodate
mismatching ions. Lithium again is favored, because it is
apparently easier for the coordinating oxygens to approach than it
is to recede from a site. But generally, both types of ions also
visit mismatching sites, suggesting also some degree of
interaction between the different species.

The data on two--ion correlations shows in fact that the dynamics
of the different species in the mixed glass are not independent.
The strongest correlations were observed for ions taking up sites
just vacated by like ions, but foreign sites are also reached.
Potassium ions will jump back with increased probability, but
lithium ions are as likely to move on into a third place as from a
matching site. For them, the correlations with the different ions
should thus in fact facilitate transport.

As mentioned above the main difference between lithium and
potassium is the ability of lithium to enter a potassium site
without an immediate backjump, helped by the local oxygen
environment. The reverse is strongly suppressed. This asymmetry is
consistent with the larger mismatch evident in the residence times
for potassium. For the present case of identical lithium and
potassium concentration the transport is mainly via the matched
sites for both species. In the limit where one species is very
dilute the relevance of transport via mixed or mismatched sites
may become important and may be relevant in particular for the
smaller ion.

It may be interesting to compare the "first-principle"
observations of this simulation work with the assumptions, used in
the most recent version of the dynamic structure model
\cite{bunde:2004b,ingram:2004} for the MA effect in solid glasses.
Most ingredients like the presence of two different types of sites
or the asymmetry between the smaller and the larger cation fully
agree with our observations. There are, however, two aspects which
are somewhat more speculative and where the present simulations
yield new insight. Do sites change their character on time scales
much shorter than the $\alpha$ relaxation time? Do so-called
C$^\prime$ sites exist? They are viewed as sites which after being
vacated in the melt are somewhat smaller than the sites, actually
visited by ions. From the simulations we have seen that on time
scales up to the diffusive regime sites keep their identity so
that the notion of a dynamic energy landscape is not supported by
the simulation data. Furthermore there is no signature of possible
C$^\prime$ sites. They would be relevant if on the diffusion time
scale transformations from C$^\prime$ sites to regular sites and
vice versa would occur. This would show up in a large number of
sites, visited by the ions during the simulation run. Since the
total number of sites, however, is just 8\% larger than the number
of ions there is basically no place for a large number of
C$^\prime$ sites being transformed to regular sites and vice
versa.

In summary, a MA slowdown can be observed in the residence times
as well as from increased likelihood of backjumps. Independent
jump paths are favored. But the interception is not complete, as
even mismatching sites are also accessible. Although actual
cooperativity seems limited to the slowest ions, an enhancement of
cation mobility is likely to follow. Together with the fast
network dynamics in MA systems this rationalizes the enhanced
dynamics in MA systems as compared to the predicted dynamics by
assuming a dilution effect. Furthermore a significant asymmetry in
the behavior of the smaller and the larger cation is observed. It
will be interesting to see, whether an even closer analysis of the
MA system, e.g. in terms of site and saddle energies, will be
compatible with very recent explicit models of the dynamics in MA
systems \cite{maass:2005}. Furthermore a closer relation to
ion-exchange experiments might be illuminating.


\begin{acknowledgments}
H. L. acknowledges funding by the german BMBF through the Fonds
der Chemischen Industrie. This work was also supported by the DFG
(SFB 458) . Furthermore we would like to thank A. Bunde, J.
Horbach, M.D. Ingram, P. Maass, and H. Mehrer for stimulating
discussions about this work.
\end{acknowledgments}


\end{document}